%% file: main.tex
\documentclass[10pt, conference, letterpaper]{IEEEtran}
\usepackage[bottom= 1.0382in,top=0.75in, left=0.6801in, right = 0.6801in]{geometry}
\usepackage{cite}
\usepackage{amsmath,amssymb,amsfonts}
\usepackage{dsfont}
\usepackage[linesnumbered,ruled,vlined]{algorithm2e}
\usepackage{graphicx}
\usepackage{textcomp}   
\usepackage{multirow}
\usepackage{setspace}

\usepackage{xcolor}
\usepackage[font=small]{caption}
\usepackage{subcaption}
\usepackage{comment}
\usepackage{url}
\usepackage{amsthm}

\usepackage{pdfpages}
\newcommand{\iq}{\mathtt{IQ}}
\newcommand{\spec}{\mathtt{S}}
\newcommand{\radar}{\mathtt{RADAR}}
\newcommand{\noradar}{\mathtt{NO\_RADAR}}
\newcommand{\fiveg}{\mathtt{5G}}
\newcommand{\detection}{\mathtt{D}}
\newcommand{\identification}{\mathtt{I}}
\newcommand{\scores}{\mathbf{s}}
\newcommand{\totalradar}{N}

    \usepackage[normalem]{ulem} 

\newtheorem{observation}{Observation}

\def\BibTeX{{\rm B\kern-.05em{\sc i\kern-.025em b}\kern-.08em
    T\kern-.1667em\lower.7ex\hbox{E}\kern-.125emX}}
\begin{document}
\title{ Pushing the Boundaries in CBRS Band: Robust Radar Detection within High 5G Interference}


\author{\IEEEauthorblockN{
         Shafi Ullah Khan\IEEEauthorrefmark{1}, 
         Michel Kulhandjian\IEEEauthorrefmark{2},
         Debashri Roy\IEEEauthorrefmark{1}
    } 
     \IEEEauthorblockA{
     \small
         \IEEEauthorrefmark{1} The University of Texas at Arlington,
         \IEEEauthorrefmark{2} Rice University\\
         Emails: \footnotesize{{shafiullah.khan@uta.edu, michel.kulhandjian@rice.edu, debashri.roy@uta.edu}}
        \vspace{-24pt} } }

\maketitle

\setstretch{0.93}

\begin{abstract}
\input{sections/abstract}

\end{abstract}
 

\vspace{-0.1in} 
\input{sections/Introduction}
\vspace{-0.1in} 
\input{sections/Related_work_full}
\vspace{-0.1in} 
\input{sections/System_model_full}

\vspace{-0.1in} 
\input{sections/Proposed_method_full}
\vspace{-0.1in} 
\input{sections/Simulation_full}
\vspace{-0.1in} 
\input{sections/Conclusion}

\bibliographystyle{IEEEtran.bst}
\bibliography{reference}

\end{document}

%% file: sections/abstract.tex
Spectrum sharing is a critical strategy for meeting escalating user demands via commercial wireless services, yet its effective regulation and technological enablement, particularly concerning coexistence with incumbent systems, remain significant challenges. Federal organizations have established regulatory frameworks to manage shared commercial use alongside mission-critical operations, such as military communications. This paper investigates the potential of machine learning (ML)-based approaches to enhance spectrum sharing capabilities within the Citizens Broadband Radio Service (CBRS) band, specifically focusing on the coexistence of commercial signals (e.g., 5G) and military radar systems. We demonstrate that ML techniques can potentially extend the Federal Communications Commission (FCC)-recommended signal-to-interference-plus-noise ratio (SINR) boundaries by improving radar detection and waveform identification in high-interference environments. Through rigorous evaluation using both synthetic and real-world signals, our findings indicate that proposed ML models, utilizing In-phase/Quadrature (IQ) data and spectrograms, can achieve the FCC-recommended $99\%$ radar detection accuracy even when subjected to high interference from 5G signals upto -5dB  SINR, exceeding the required limits of $20$ SINR.  Our experimental studies distinguish this work from the state-of-the-art by significantly extending the  SINR limit for $99\%$ radar detection accuracy from approximately $12$ dB down to $-5$ dB. Subsequent to detection, we further apply ML to analyze and identify radar waveforms. The proposed models also demonstrate the capability to classify six distinct radar waveform types with $93\%$ accuracy.

%% file: sections/Introduction.tex
\section{Introduction}
\label{sec:introduction}

\begin{figure}[t]
    \centering
    \includegraphics[width=1\linewidth]{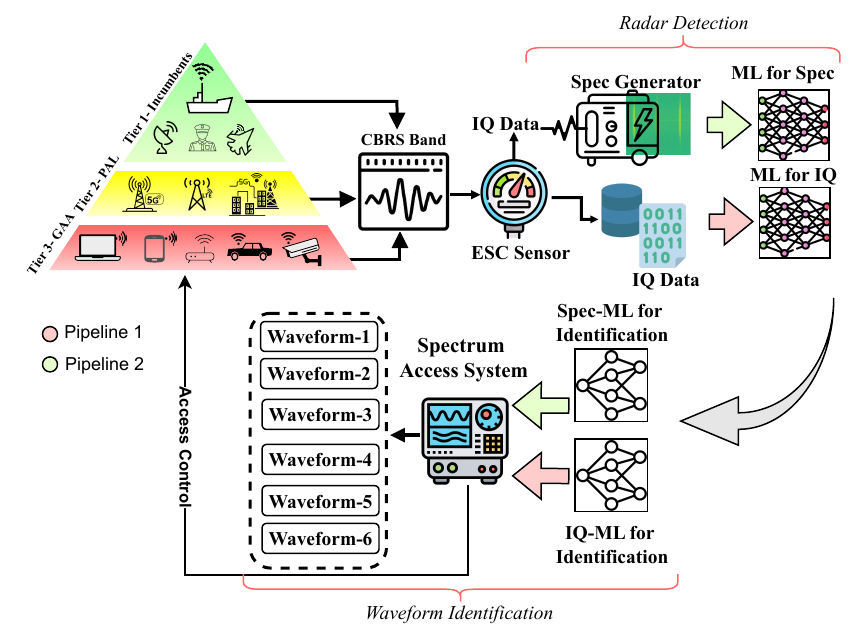}
    \vspace{-20pt}
    \caption{Proposed machine learning pipelines for radar detection in the presence of $5$G interference and subsequent radar waveform identification. }
    \label{fig:intro_fig}
    \vspace{-20pt}
\end{figure}

\noindent {\bf Regulations for the Shared Spectrum Usage.} The exponential growth in wireless connectivity and the emergence of data-intensive applications have significantly intensified the demand for radio spectrum \cite{8255748}. To address this challenge, modern spectrum management strategies are shifting from static allocation to dynamic spectrum sharing paradigms. A prominent example is the Citizens Broadband Radio Service (CBRS) band, spanning $3.55–3.7$ GHz, which was authorized by the U.S. Federal Communications Commission (FCC) to facilitate shared access among diverse users while safeguarding incumbent federal operations, primarily naval radar systems\cite{FCC_CBRSpage}. Under this framework, users are categorized into a three-tier hierarchy as shown in Fig. \ref{fig:intro_fig}. Tier $1$ for incumbents, Tier $2$ for Priority Access Licensees (PAL), and Tier $3$ for General Authorized Access (GAA) users~\cite{Shafi2025}. Spectrum access is managed by a central entity known as Spectrum Access System (SAS) that has no control over the incumbent users, but grants access to the registered PAL and GAA users~\cite{FCC1547}. Both PAL and GAA users need to register with SAS to be considered as authorized users, and to be granted access based on their priority. The SAS relies on sensing information from an array of geographically distributed Environmental Sensing Capability (ESC) sensors for allocating the spectrum to PAL and GAA users. 

While shared access improves spectral efficiency, it introduces critical challenges, particularly with regard to interference between coexisting systems. Radar signals, as primary incumbents, must be reliably detected and protected from harmful interference generated by secondary users.  Hence, for a trusted shared spectrum usage, FCC has mandated that ESCs should be designed such that detection of $99\%$ of radar pulses in signal-to-interference-plus-noise ratio (SINR) $20$\,dB is guaranteed~\cite{lab-radar}. Commercial wireless service providers operating $4$G, $5$G, and private networks typically fall under the PAL and GAA categories, subject to strict transmit power constraints. Hence, with radar pulses having at least $-89$\,dBm/MHz power, the cumulative power of additive white Gaussian noise (AWGN) and other sources of interference must be no more than $-109$\,dBm/MHz around the ESC sensor~\cite{CBRS-Requirement}. 

\noindent {\bf Pushing the Boundaries.} 
In this paper, we push this boundary by a implementing a Machine Learning (ML)–based framework for ESC sensor that yields to $99\%$ detection capability at SINR levels as low as $-5$\,dB, exceeding the FCC-mandate by $25$\,dB. In other words, if the power of radar pulses are $-94$\,dBm/MHz, our method will be able to provide FCC-mandated radar detection accuracy of $99\%$ even when the noise plus interference power is as high as $-89$\,dBm/MHz. This sets us apart from the state-of-the-art, where the ML-based approaches achieve the FCC-mandated $99\%$ radar detection accuracy upto SINR of $20$\,dB~\cite{DeepRadar}, $17$\,dB~\cite{10001638}, $16$\,dB~\cite{10804612}, $15$\,dB~\cite{10976013}, and $12$\,dB~\cite{senseORAN}, hence maximum allowable  noise plus interference power would be: $-114$\,dBm/MHz, $-111$\,dBm/MHz, $-110$\,dBm/MHz, $-109$\,dBm/MHz, $-106$\,dBm/MHz, respectively.

Additionally, our framework identifies the radar waveforms, thereby providing spectrum-access systems with precise information on the active incumbent signal and enabling more informed, interference-aware spectrum management.

\noindent {\bf Novel Contributions.} As shown in Fig.~\ref{fig:intro_fig}, we propose two pipelines: (a) {\tt Pipeline 1} and (b) {\tt Pipeline 2}. The {\tt Pipeline 1}  detects the presence of radar signal and identify its waveform from the IQ data. As IQ data expose the  phase information of signal, hence, prone to security vulnerabilities for sharing~\cite{10824882}. So, we propose {\tt Pipeline 2}, which does similar radar detection and waveform identification, but from spectrograms with extra processing time for spectrogram generation time. Overall, our main contributions are summarized as follows:

\begin{itemize}
\item We propose a ML-based solution for detecting radar from In-phase Quadrature (IQ) data within high interference 5G signals with SINR of $\geq-5$\,dB ({\tt Pipeline 1}). 

\item We further propose a ML-based solution for radar detection with SINR of $\geq-5$\,dB  using spectrogram ({\tt Pipeline 2}). 

\item We also propose a hierarchical classifier which does radar waveform classification, enabling fine-grained identification of radar waveforms beyond mere presence detection.

\item We simulate a diverse set of radar waveforms that emulate chirp-based radar signals, generating a comprehensive dataset comprising six distinct radar waveforms with varied signal parameters. 

\item Finally, we validate the proposed approach using over-the-air data collected via a custom software defined radio (SDR)-based testbed in a controlled laboratory environment. The framework achieves over $99\%$ detection accuracy on real-world signals, demonstrating strong generalization capability and compliance with regulatory standards. We release our code and data for community usage at \cite{TWISTLab_RadarPushingBound_GitHub_2025}.
\end{itemize}

%% file: sections/Related_work_full.tex
\section{Related Work}
\label{sec:related_work}
Conventional radar-sensing systems in shared-spectrum environments rely on energy detectors. Recently, there have been various ML-based efforts, such as, Lees \textit{et al.} demonstrated the superiority of deep learning techniques over traditional radar detection algorithms by generating independent spectrograms for each $10$~MHz channel and utilizing them for radar signal detection~\cite{Lees_2019}. Along similar lines, Basak \textit{et al.} trained a variant of the YOLO-lite framework to classify ten signal classes, including two Wi-Fi versions and eight drone signals in the Industrial, Scientific, and Medical (ISM) band~\cite{9493767}. Their study utilized over-the-air datasets containing single or multiple overlapping signals in both isolated and coexistent scenarios. Furthermore, the work in~\cite{9677280} employed multiple convolutional neural network (CNN) architectures for radar detection in the CBRS band, exploring both raw magnitude-based and spectrogram-based binary classification approaches to determine the presence of radar signals. Building on this direction, the ESC+ framework introduced in~\cite{10001638} leveraged spectrograms and YOLO for radar detection in the CBRS band, achieving over $99\%$ accuracy at SINR levels as high as $17$\,dB. Additionally, the Spec-SCAN system proposed in~\cite{10976013} adopted a localized scanning strategy using spectrogram analysis for confined frequency bands, reaching a recall of $99\%$ for a single radar class at a minimum SINR of $15$\,dB. SenseORAN \cite{senseORAN} embeds a modified YOLO-based xApp within the gNB’s near-RT RIC to detect radar signals in the $3.5$\,GHz CBRS band by converting uplink IQ samples into spectrograms and maintaining a dynamic occupied-channel list; by fusing seven spectrograms per inference, it achieves 100\% detection at SINR$\geq$12\,dB.

\noindent {\bf Innovation Opportunity.} While these works showcase the effectiveness of spectrogram-based deep learning approaches, to the best of our knowledge, no existing study has demonstrated robust radar detection at SINR levels lower than $12$\,dB. Moreover, prior literature primarily focuses on binary detection and lacks the capability to distinguish between various radar waveforms coexisting with 5G signals in the CBRS band.


%% file: sections/System_model_full.tex
\section{Problem and Solution}
\label{sec:system_model}

\vspace{-4pt}
\subsection{Radar Interference in 5G Wireless System}
\label{sec:nbiot}
\vspace{-2pt}
\subsubsection{Radar Specifications}
In our system, we consider a typical airborne scenario with radar transmitting an linear frequency modulated pulse (LFM) waveform. We represent the LFM waveforms by:\vspace{-1mm}
\begin{equation} \label{eq:signalModel}
    s(t) = \sum_{p=0}^{P-1}  x(t - pT) \Pi\left ( \frac{t - pT}{\Omega} \right ),
\end{equation}
where $P$ is the number of pulses, and $T$ is the pulse repetition interval, $1/T$ pulse repetition frequency (PRF), $\Pi (\cdot)$ is defined as a rectangular 
with a pulse duration/width (PW) of $\Omega$ and the component pulse $x(.)$ is given as
\begin{equation}
    x(t) = A e^{j\left ( \phi_0 + 2\pi f_0 t  + \pi g  t^2 \right )},
\end{equation}
\noindent where $A$ is the amplitude of the pulse, $\phi_0$ is an initial phase offset, $f_0$ is the starting frequency, and $g$ is the \textit{chirp rate} defined as $g = B/\Omega$ Hz/s, where $B$ is the pulse bandwidth (BW). 

The variation of these parameters generate different radar waveforms. We consider $\totalradar$ type of such radar pulse waveforms: $\mathcal{W} = \{ Waveform\,1, \cdots, Waveform\,N \} $, where $|\mathcal{W}| = \totalradar$.
\vspace{-4pt}
\subsection{Problem Formulation}
We consider a shared spectrum environment in which incumbent radar systems coexist with commercial 5G communication systems within the CBRS band. Let $s(t)$ denote the radar signal and $p(t)$ denote the 5G signal. The signal received at the ESC sensor is modeled as,
\begin{equation}
r(t) =  \alpha * s(t - \tau) + \mu^{\text{CR}} * p(t) + n(t),
\end{equation}
where  `$*$' signify multiplication, $\alpha$ is the backscattering coefficient of the radar echo, $\tau$ is the propagation delay, $\mu^{\text{CR}}$ is the interference from the transmitter, and $n(t)$ is additive noise.

We define three signal coexistence scenarios:
\begin{itemize}
    \item \textbf{Scenario $1$ (Radar-only):} $p(t) = 0$. The ESC must detect radar activity and classify the waveform type.
    \item \textbf{Scenario $2$ (5G-only):} $s(t) = 0$. The ESC must confirm radar absence to allow spectrum access.
    \item \textbf{Scenario $3$ (Radar + 5G):} Both $s(t) \neq 0$ and $p(t) \neq 0$. The ESC must reliably detect radar presence despite interference and further identify radar waveform from a known set. The objective is to design a decision function $\mathcal{D} (\cdot)$ that maps received signal $r(t)$ to a binary decision:
\end{itemize}

\vspace{-10pt}
\begin{equation}
\mathcal{D} (r(t)) \in \{ \radar, \noradar\}.
\vspace{-3pt}
\end{equation}

In the case where $\mathcal{D}(r(t)) = \radar$, a hierarchical classifier $\mathcal{I}(\cdot)$ further identify $\totalradar$ different radar waveforms:
\vspace{-3pt}
\begin{equation}
\mathcal{I}(r(t)) \rightarrow \text{Radar Waveform} \in  \mathcal{W}.
\end{equation}

%% file: sections/Proposed_method_full.tex
\vspace{-4pt}
\subsection{Proposed Solution: Radar Detection}
\label{sec:proposed_work_detection}
\vspace{-4pt}
\noindent{\bf \texttt{Pipeline 1:}} The training data matrix for IQ network is composed of the IQ samples $\iq_{\fiveg}$, $\iq_{\radar}$ and $\iq_{\fiveg+\radar}$ from  $5$G, radar  and overlapped $\radar$+$\fiveg$ signals, respectively. To adapt to our binary classification problem, our radar data matrix $X_{\iq^{\radar}}$  consists of $\iq_{\radar}$ and $\iq_{\fiveg+\radar}$ IQ samples and without radar data matrix $X_{\iq^{\noradar}}$ consists of $\iq_{\fiveg}$ IQ samples. Overall our training data matrix is denoted as: $X_{\iq}\in \mathbb{R}^{ d_0^\iq \times 2} = \{X_{\iq^{\radar}}, X_{\iq^{\noradar}}\}$, where $(d_0^\iq \times 2)$ is the dimensionality of the IQ samples with $2$ representing the I and Q components. The set of the output labels are: $\mathcal{L}_{\detection} = \{\radar, \noradar\}$. We consider the label matrix $Y_{\detection}\in \{0,1\}^{ |\mathcal{L}_{\detection}|}$ that represent the one-hot encoding for either radar present or absent.  The prediction vector $\scores^{\detection}_{\iq}\in \mathbb{R}^{ |Y_{\detection}|}$  is generated through a Softmax activation $\sigma$, as:
\vspace*{-0.10in}
\begin{equation}
\label{eq:fusion_network_anomaly}%
 \vspace*{-0.05in}
 \scores^{\detection}_{\iq}=
\sigma(f_{{\theta}_{\detection}^{\iq}}^{\iq}(X_{\iq})), \;f_{{\theta}_{\detection}^{\iq}}^{\iq}: \mathbb{R}^{N^{d_0^\iq \times 2}} \mapsto  \mathbb{R}^{|Y_{\detection}|} .
\end{equation}

 Overall the radar detection problem is solved using the IQ network by: $ \mathcal{D}  (.) = \arg\max \sigma(f_{{\theta}_{\detection}^{\iq}}^{\iq}(.))$. 

\noindent{\bf \texttt{Pipeline 2:}} 
In this case, the spectrogram samples are denoted as: $\spec_{\radar}$, $\spec_{\fiveg}$ and $\spec_{\radar+\fiveg}$ from spectrograms of radar, $5$G,  and overlappped $\radar$+$\fiveg$ signals, respectively. This data matrix is denoted as: $X_{\spec}\in \mathbb{R}^{d_0^{\spec} \times d_1^{\spec} \times 3} = \{X_{\spec^{\radar}}, X_{\spec^{\noradar}}\}$, with $ X_{\spec^{\radar}}$ consisting of $\spec_{\radar}$ and $\spec_{\noradar}$ spectrogram samples and $X_{\spec^{\noradar}}$ consisting of $\spec_{\fiveg}$ samples. The $(d_0^{\spec} \times d_1^{\spec} \times 3)$ gives the dimensionality of spectrogram images with 3 representing the three channels for RGB. The labels and label matrix are same as IQ network. The prediction vector $\scores^{\detection}_{\spec}\in \mathbb{R}^{|Y_{\detection}|}$ through a Softmax activation $\sigma$, is denoted as:
 \vspace*{-0.10in}
\begin{equation}
\label{eq:fusion_network_anomaly}%
 c
 \scores^{\detection}_{\spec}=
\sigma(f_{{\theta}_{\detection}^{\spec}}^{\spec}(X_{\spec})), \;f_{{\theta}_{\detection}^{\spec}}^{\spec}: \mathbb{R}^{N^{d_0^\spec \times d_1^\spec\times 3}} \mapsto  \mathbb{R}^{|Y_{\detection}|} 
\end{equation}

In this case, $ \mathcal{D}  (.) = \arg\max \sigma(f_{{\theta}_{\detection}^{\spec}}^{\spec}(.))$.
\vspace{-10pt}
\subsection{Proposed Solution: Radar Waveform Identification}
\label{sec:proposed_work_identification}

After radar is detected within a signal, we propose the second stage of hierarchical classifier to detect the waveform of the radar signal out of $\totalradar$ possibilities. 

\noindent{\bf \texttt{Pipeline 1:}}
In this case, the data matrix consists of IQ samples with radar signals, hence $X_{\iq^{\radar}}\in \mathbb{R}^{ d_0^\iq \times 2}$. The label matrix is $Y_{\identification}\in \{0,1\}^{ \totalradar}$ in the one-hot encoding representation to signify $\totalradar$ possibilities of radar waveforms. The prediction vector $\scores^{\identification}_{\iq}\in \mathbb{R}^{|Y_{\identification}|}$ through a Softmax activation $\sigma$, is denoted as:
\vspace{-10pt}
\begin{equation}
\label{eq:fusion_network_anomaly}%
 \vspace*{-0.05in}
\scores^{\identification}_{\iq}=
\sigma(f_{{\theta}_{\identification}^{\iq}}^{\iq}(X_{\iq^{\radar}})), \;f_{{\theta}_{\identification}^{\iq}}^{\iq}: \mathbb{R}^{N^{d_0^\iq \times 2}} \mapsto  \mathbb{R}^{|Y_{\identification}|}  
\end{equation}

We use IQ Network as: $ \mathcal{I}  (.) = \arg\max \sigma(f_{{\theta}_{\identification}^{\iq}}^{\iq}(.))$. to solve the radar waveform identification problem.

\noindent{\bf \texttt{Pipeline 2:}} Similar to the   \texttt{Pipeline 1}, the data matrix and label matrices are:  $X_{\spec^{\radar}}\in \mathbb{R}^{d_0^{\spec} \times d_1^{\spec} \times 3}$ and $Y_{\identification}\in \{0,1\}^{ \totalradar}$, respectively. The prediction vector $\scores^{\identification}_{\spec}\in  \mathbb{R}^{|Y_{\identification}|}$ is:
\begin{equation}
\label{eq:fusion_network_anomaly}%
 \vspace*{-0.05in}
\scores^{\identification}_{\spec}=
\sigma(f_{{\theta}_{\identification}^{\spec}}^{\spec}(X_{\spec^{\radar}})), \;f_{{\theta}_{\identification}^{\spec}}^{\spec}: \mathbb{R}^{N^{d_0^\spec \times 2}} \mapsto  \mathbb{R}^{|Y_{\identification}|}  
\end{equation}

In this case, $ \mathcal{I}  (.) = \arg\max \sigma(f_{{\theta}_{\identification}^{\spec}}^{\spec}(.))$.

%% file: sections/Simulation_full.tex
\vspace{0.1cm}
\section{Experiments}
\label{sec:results}
To validate our proposed pipelines for both radar detection and waveform identification, we generate synthetic dataset with varies SINR and real-world dataset from a SDR testbed. 
\subsection{Synthetic Dataset}
We generate a synthetic IQ dataset using MATLAB's Communication, Array Processing, and $5$G Toolboxes. The simulation involves creating a $5$G physical layer signal by placing a base station, multiple user equipments, and a radar system at specific $3$D locations. The dataset features a radar signal with different SINRs relative to $5$G and noise levels, sampled at $61.44$ MHz with a center frequency of $3.6$ GHz, modeling the interaction between radar and $5$G signals in a realistic shared spectrum environment.  We define SINR as $10 \times \log_{10}(P_{\text{radar}}/P_{\text{noise}})$, where $P_{\text{radar}}$ is peak radar power per MHz, and $P_{\text{noise}}$ is average noise plus interference power per MHz of the radar band. 
{\color{black}We vary the power of radar pulses in the range of $-89$ to $-79$\,dBm/MHz to emulate SINRs in the range of $-20$ to $20$\,dB, with $5$\,dB interval. For the noise and interference power, we go upto $25$ dB beyond FCC limit, and allow the noise and interference power to vary in the range of $-109$ to
$-84$\,dBm/MHz. Overall, our synthetic dataset contains $\sim 13608$ generated signals from three categories: (a) radar pulses only, (b) 5G signals only, and (c) radar pulses overlapped with 5G signals, out of which we utilize $4536$ samples with SINR varying between $-5$\,dB to $5$\,dB for this work.}
We consider six distinct LFM chirp waveforms, representative of common radar types used in tactical and military communications.  Table~\ref{tab:waveforms}, shows a wide range of radar system characteristics. BIN$1$-A and BIN$1$-B represent short-duration, high-PRF radars, that are suitable for high-resolution target tracking or missile guidance systems. Their high chirp rates (up to $4$~GHz/s) make them ideal for fine range discrimination. BIN$2$ and BIN$3$ waveforms correspond to medium- and long-pulse systems, often used in surveillance, marine, or ground-based tracking radars.

\begin{table}[ht]
\centering
\vspace{-4pt}
\caption{Specifications of various radar waveforms.}
\label{tab:waveforms}
\resizebox{0.45\textwidth}{!}{%
\begin{tabular}{|l||c|c|c|c|}
\hline
\textbf{Waveform} & \textbf{PW ($\mu$s)} & \textbf{PRF (Hz)} & \textbf{BW (MHz)} & \textbf{Chirp Rate (GHz/s)} \\
\hline
\hline
BIN1-A & $1$   & $20,000$ & $1$  & $1000$ \\
BIN1-B & $2$   & $30,000$ & $8$  & $4000$ \\
BIN2-A & $50$  & $200$    & $20$ & $400$  \\
BIN2-B & $100$ & $200$    & $30$ & $300$  \\
BIN3-A & $10$  & $100$    & $5$  & $500$  \\
BIN3-B & $50$  & $150$    & $15$ & $300$  \\
\hline
\end{tabular}%
 \vspace{-0.2in}
}
\end{table}
\vspace{-3pt}

\subsection{Real-world Dataset}
\begin{figure}[!hbt]
    \centering
    \vspace{-10pt}
    \includegraphics[width=0.75\linewidth]{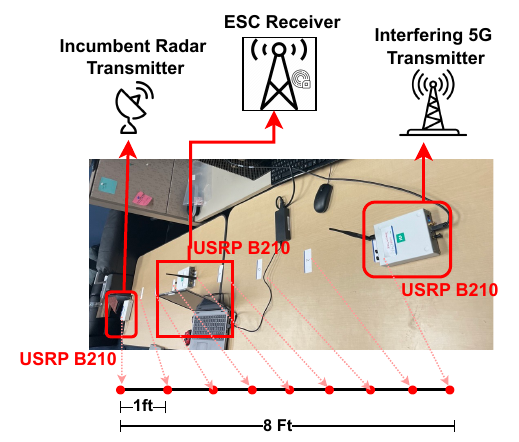}
    \caption{Laboratory testbed emulating a real‐world spectrum‐coexistence scenario with an incumbent radar transmitter, ESC receiver, and interfering $5$G transmitter.}
    \label{fig:Laboratory setup}
     \vspace{-0.2in}
\end{figure}

Our SDR testbed consists of multiple USRP B$210$ devices in a controlled laboratory environment. Specifically, one USRP B$210$ is configured to transmit radar pulses, another to transmit $5$G waveforms, and a third to function as a receiver, capturing the over-the-air signals. The experiment incorporates six distinct radar pulse types alongside $5$G signals, transmitted under various controlled conditions. To evaluate the robustness of the framework under noisy and dynamic channel conditions, we systematically varied the distance between the transmitters and the receiver, thereby introducing controlled noise levels into the received signal. Multiple samples were collected for each configuration to ensure statistical validity. Additionally, signal diversity was introduced by varying the transmission bandwidth of the $5$G waveforms. All transmissions are carried out at a center frequency of $3.65$\,GHz with a constant sampling rate of $50$\,MS/s and a fixed antenna gain of $85$\,dB. The experimental setup is illustrated in Fig.~\ref{fig:Laboratory setup}, while the detailed parameter configurations and signal characteristics are summarized in Table~\ref{tab:signal-params}.

\begin{table}[!hbt]
    \centering
    \caption{The specification of various real world data collection scenario. We emulate various signal interference by varying the distance of the radar transmitter and $5$G transmitter from the receiver. }
    \label{tab:signal-params}
 \resizebox{0.49\textwidth}{!}{   \begin{tabular}{|l|p{0.5in}|c|c|c|c|c|}
        \hline
        \textbf{Signal Type} & 
        \shortstack{\textbf{Distance} \\ {\textbf{(Tx--Rx)}}} & 
        \textbf{BW (MHz)} & 
        \shortstack{\textbf{Samp} \\ \textbf{Rate}} & 
        \shortstack{\textbf{Center} \\ \textbf{Freq (GHz)}} & 
        \shortstack{\textbf{Samples}}  & 
        \shortstack{\textbf{Ant} \\ \textbf{Gain}} \\
        \hline \hline
        \textbf{Radar}       & $1$--$8$  ft   & $20$        & $50$ MS/s       & $3.65$       & $420$ & $85$\,dB \\
        \textbf{$5$G}          & $1$--$8$ ft    & $5$--$50$      & $50$ MS/s       & $3.65$        & $480$ & $85$\,dB \\
        \textbf{Radar + $5$G}  & $1$--$8$   ft  & $20$ + $5$--$50$    & $50$ MS/s       & $3.65$        & $480$ & $85$\,dB \\
        \hline
    \end{tabular}}
\end{table}

\noindent {\bf Dataset Availability}: Overall, the statistics of the IQ and spectrogram data are: (a) $1,512$ for each SINR $-5$\,dB  ,$0$\,dB, $5$\,dB and (b) $918$ samples for real-world data, incurring total number of $5,454$. Samples of the spectrograms and IQ data of both the synthetic and real datasets are available at \cite{TWISTLab_RadarPushingBound_GitHub_2025}.

\subsection{Neural Network (NN) Architectures}

\begin{figure}
    \centering
    \vspace{-10pt}
    \includegraphics[width=0.9\linewidth]{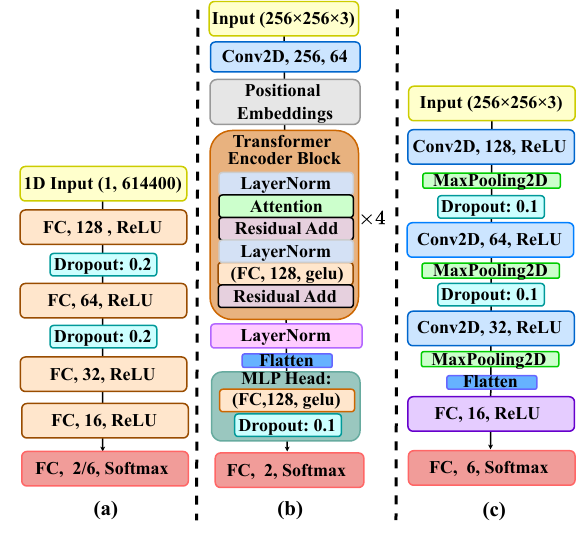}
    \caption{Different neural network architectures used for the two parts of hierarchical classifier using: (a) IQ samples for radar detection (\texttt{Pipeline 1}) and waveform identification (\texttt{Pipeline 2}), (b) Spectrograms for radar detection (\texttt{Pipeline 2}), (c) Spectrograms for radar waveform identification (\texttt{Pipeline 2}).}
    \label{fig:nn_architectures}
    \vspace{-0.25in}
\end{figure}

The details of NN models $f_{{\theta}_{\detection}^{\iq}}^{\iq}(.)$ and $f_{{\theta}_{\detection}^{\spec}}^{\spec}(.)$ (discussed in Section~\ref{sec:proposed_work_detection}) for the {\em radar detection} part of the hierarchical classifier are shown in  Fig.~\ref{fig:nn_architectures} (a) and (b). Similarly, the NN models $f_{{\theta}_{\identification}^{\iq}}^{\iq}(.)$ and $f_{{\theta}_{\identification}^{\spec}}^{\spec}(.)$ (as discussed in Section~\ref{sec:proposed_work_identification}) for the {\em radar waveform identification} part of the hierarchical classifier is shown in  Fig.~\ref{fig:nn_architectures} (a) and (c). We exploit categorical cross-entropy loss for training using Tensorflow backend with $70/20/10$ train/val/test ratio, details of other parameters for different NN model are in Table~\ref{tab:hyperparameters}.

\vspace{16pt}
\begin{table}[h!]
\centering
\vspace{4.5pt}
\caption{Hyperparameter settings for training various NN models.}
\vspace{-4pt}
\label{tab:hyperparameters}
\resizebox{0.48\textwidth}{!}{%
\begin{tabular}{|p{1in}||p{1in}|p{1in}|p{1in}|}
\hline

\textbf{Hyperparameter}         & \textbf{ViT} (Fig.~\ref{fig:nn_architectures} (b))                         & \textbf{CNN} (Fig.~\ref{fig:nn_architectures} (c))            & \textbf{ANN}  (Fig.~\ref{fig:nn_architectures} (a))            \\ \hline \hline
Batch size                      & $32$                                  & $32$                       & $16$                  \\ \hline
Input channels                  & RGB ($3$)                             & RGB ($3$)                 & --                        \\ \hline
Learning rate                   & $5\times10^{-4}$                      & $5\times10^{-4}$         & $1\times10^{-4}$          \\ \hline
Activation                      & Softmax                               & Softmax                  & Softmax                   \\ \hline
Loss                            & categorical crossentropy             & categorical crossentropy & categorical crossentropy  \\ \hline
Epochs                          & $200$                                & $100$                    & $100$                       \\ \hline
Optimizer                       & Adam                                  & Adam                     & Adam                      \\ \hline
Test split                      & $15\%$                                  & $15\%$                     & $30\%$                     \\ \hline
Validation split                & $10\%$                              & $10\%$                  & $10\%$                      \\ \hline
Dropout rate                    & $0.1$                               & --                       & --                        \\ \hline
Projection dimension \& layers  & $64$ ($4$ transformer layers)             & --                       & --                        \\ \hline
Kernel size & $16$                & $3$\,$\times$\,$3$          & --                        \\ \hline
Number of Conv layers  & $1$                                    & $4$                        & --               \\ \hline
\end{tabular}%
} 
\end{table}

\noindent {\bf Evaluation Metrics.} We use standard classification metrics: accuracy, precision, recall, F$1$ score, and inference time to validate our proposed pipelines for both radar detection and radar waveform identification task.

\subsection{Experimental Results on Radar Detection} 

In the first set of experiments, we evaluate the {\em radar detection} part of the hierarchical classifier that detects radar presence. The performance of all the pipelines across different datasets is shown  in Table~\ref{tab:radar_detection_performance}.  Overall, \texttt{Pipeline 2} gives better performance than \texttt{Pipeline 1}. Hence, we perform a trade-off analysis of these two pipelines in terms of inference time for radar detection and performance. 
\vspace{-0.2cm}

\begin{observation}

We observe that the spectrogram-based \texttt{Pipeline 2} maintain $\geq$99\% accuracy even at $–5$\,dB SINR, while the IQ-based \texttt{Pipeline 1} fails to achieve FCC mandated $99\%$ radar detection accuracy (refer to Table~\ref{tab:radar_detection_performance}).
\end{observation}

\begin{table}[h!]
\centering
\vspace{5pt}
\caption{Performance of {\tt Pipeline 1} and {\tt 2} for radar detection. }
\label{tab:radar_detection_performance}
\resizebox{0.43\textwidth}{!}{\begin{tabular}{|c|c||c|c|} 
\hline
\multirow{2}{*}{\textbf{Signal}} & \multirow{2}{*}{\textbf{Metrics ($\%$)}} & {\textbf{Spectrograms}} & \textbf{IQ Samples}  \\ 
\cline{3-4}
                                 &        & {\tt Pipeline 2}    & {\tt Pipeline 1}     \\ 
\hline
\hline
\multirow{4}{*}{SINR $5$\,dB}          & Overall Accuracy      & $99.8$  & $98.02$  \\
                                 & Radar Precision        & $99.6$  & $100$    \\
                                 & False Positives              & $0$     & $0$      \\
                                 & Recall                & $99.5$  & $94.17$  \\ 
\hline
\multirow{4}{*}{SINR $0$\,dB}          & Overall Accuracy      & $99.7$  & $97.64$  \\
                                 & Radar Precision        & $99.6$  & $99.5$   \\
                                 & False Positives              & $0$     & $0$      \\
                                 & Recall                & $99.5$  & $93.27$  \\ 
\hline
\multirow{4}{*}{SINR $-5$\,dB}   & Overall Accuracy       & $99.5$  & $96.54$  \\
                                 & Radar Precision        & $99.5$  & $98.25$  \\
                                 & False Positives              & $0$     & $0$      \\
                                 & Recall                 & $99.5$  & $91.17$  \\ 
\hline
\multirow{4}{*}{Real Data}       & Overall Accuracy     & $99.5$  & $91.10$  \\
                                 & Radar Precision       & $99.6$  & $97.40$  \\
                                 & False Positives               & $0$     & $0$      \\
                                 & Recall                 & $99.5$  & $88.20$  \\ 
\hline
\end{tabular}}
\vspace{-15pt}
\end{table}

\begin{figure}[!hbt]
    \centering
    \vspace{-15pt}
    \includegraphics[width=0.7\linewidth]{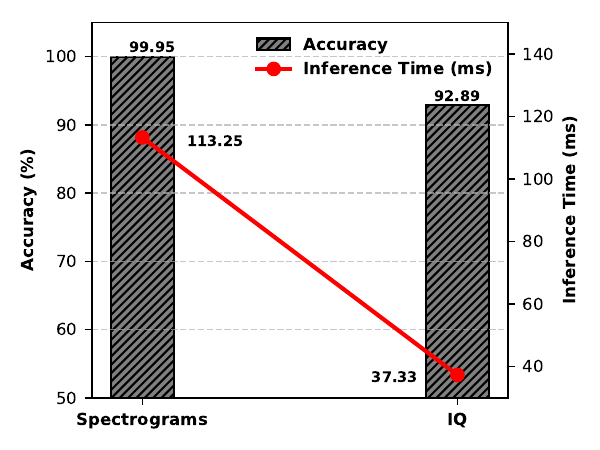}
    \caption{Trade-off between the inference time and the radar detection accuracy at SINR= $-5$dB. }
    \label{fig:inference}
    \vspace{-10pt}
\end{figure}

\noindent{\bf Trade‐off Analysis of {\tt Pipeline 1} vs. {\tt Pipeline 2}.}
Fig.~\ref{fig:inference} illustrates the inference time radar detection vs. accuracy trade-off between the spectrogram-based {\tt Pipeline 2} and IQ-based  {\tt Pipeline 1}. The {\tt Pipeline 2} applies a Short-time Fourier Transform (STFT) to the $20$\,ms of IQ values sampled at $61.44$\,MHz, incurring approximately $80$\,ms to generate the time–frequency spectrogram followed by $33$\,ms for model inference (total $113$\,ms). Whereas the {\tt Pipeline 1} completes end-to-end inference in $37$\,ms. Despite the additional 76\,ms preprocessing overhead, the {\tt Pipeline 2}  achieves $99.95\%$ accuracy, which is $8\%$ higher than that of the {\tt Pipeline 1}'s  $92.89\%$, by leveraging detailed time–frequency features that bolster robustness under severe interference.

\vspace{-0.2cm}
\begin{observation}
We observed that despite an approximately threefold increase in end-to-end inference latency ($113$\,ms vs.\ $37$\,ms), the spectrogram‐based model delivers an $8\,\%$ higher accuracy ($99.95\,\%$ vs. $92.89\,\%$), highlighting the latency–accuracy trade‐off inherent in time–frequency processing (refer Fig. \ref{fig:inference}).

\end{observation}

\begin{figure}[htbp]
    \centering
    \begin{subfigure}[b]{0.23\textwidth}
        \centering
        \includegraphics[width=\textwidth]{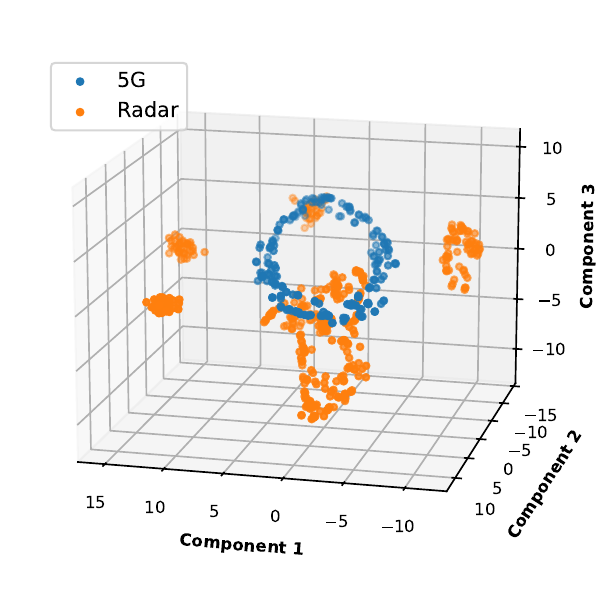}
        \caption{IQ-based {\tt Pipeline 1}}
        \label{fig:tsne_iq}
    \end{subfigure}
    \hfill
    \begin{subfigure}[b]{0.23\textwidth}
        \centering
        \includegraphics[width=\textwidth]{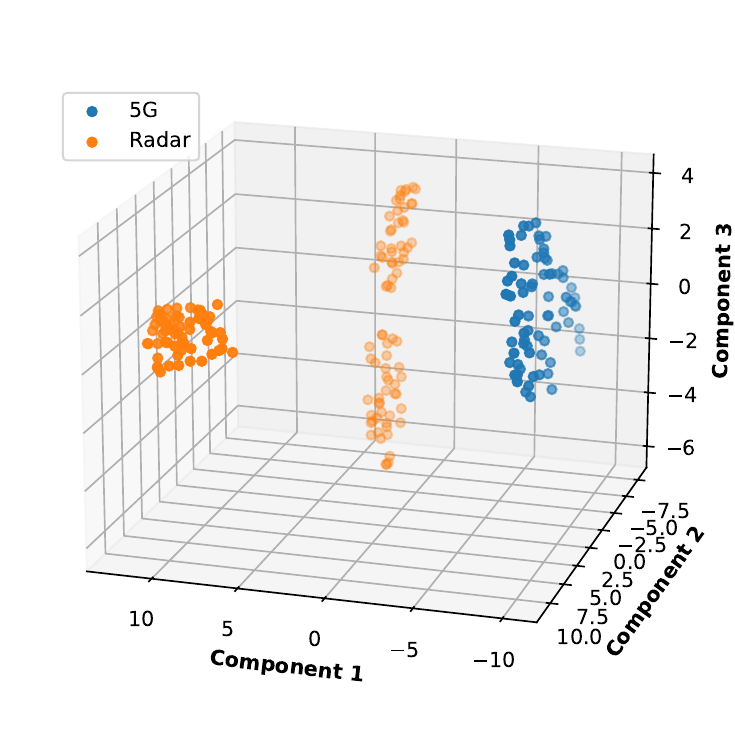}
        \caption{Spec-based {\tt Pipeline 2}}
        \label{fig:tsne_spec}
    \end{subfigure}
    \caption{Comparison of feature separability via $3$D t-SNE for IQ-based {\tt Pipeline 1} and spectrogram-based {\tt Pipeline 2}. Spectrogram features exhibit better class distinction.}
    \label{fig:tsne_comparison}
\end{figure}

\noindent{\bf t-SNE-Based Feature Space Visualization for {\tt Pipeline 1} vs. {\tt Pipeline 2}.}
t-distributed stochastic neighbor embedding (t-SNE) is a nonlinear dimensionality reduction technique used to visualize high-dimensional data in low-dimensional space while preserving local structure \cite{ZAMAN2025109787}. Figure~\ref{fig:tsne_comparison} illustrates the $3$D t-SNE projections of the learned feature spaces from the NN models of both the pipelines. As shown in Fig.~\ref{fig:tsne_iq}, the NN model of {\tt Pipeline 1} produces less distinct clusters, with noticeable overlap between the $\radar$ and $\noradar$ classes. This suggests that raw IQ inputs lack sufficient discriminative characteristics needed for precise classification in congested spectral environments. Conversely, the NN model of {\tt Pipeline 2}  in Fig.~\ref{fig:tsne_spec} achieves well-separated clusters for all three classes, indicating superior representational capability. The richer time–frequency structure captured in spectrograms allows the NN model to extract more salient features, enabling enhanced class separability.

\vspace{-14pt}

\vspace{-5pt}
\noindent{\bf Comparison with state-of-the-art.}
Table~\ref{tab:radar_detection_comparison} presents a comparative analysis between the proposed framework and several state-of-the-art radar detection methods developed for the CBRS band. While existing approaches such as RadYOLOLet~\cite{10804612}, Waldo~\cite{10001638}, DeepRadar~\cite{DeepRadar}, Spec-SCAN~\cite{10976013} and SenseORAN~\cite{senseORAN} achieve high detection accuracy, they are primarily constrained to operating at SINR levels no lower than 12\,dB. In contrast, our method achieves $100\%$ recall even in harsh interference environments, operating reliably at an SINR as low as -5\,dB, thus significantly surpassing the SINR tolerance of previous work.

In addition to robust detection, the proposed framework also performs fine-grained radar waveform classification, identifying six distinct types of radar. This level of granularity exceeds that of DeepRadar and RadYOLOLet, which classify up to five radar types, while Waldo, Spec-SCAN and SenseORAN only considered one type of identification. The combined ability to detect radar at lower SINR levels and identify more radar types highlights the enhanced sensitivity and utility of our system for spectrum co-existence and dynamic spectrum access enforcement.
\vspace{-0.2cm}
\begin{observation}
 We observe that our framework extends reliable detection down to $–5$\,dB SINR while classifying six distinct radar waveforms, surpassing prior CBRS detectors in both interference robustness and identification granularity (See Table~\ref{tab:radar_detection_comparison}).
\end{observation}

\begin{table}[h!]
\centering
\vspace{-10pt}
\caption{Comparison of state-of-the-art radar detection methods vs. proposed approach ({\tt Pipeline 2}).}
\label{tab:radar_detection_comparison}
\resizebox{0.4\textwidth}{!}{%
\begin{tabular}{|l||c|c|c|c|c|}
\hline
\textbf{Paper} &\textbf{SINR} & \textbf{Recall} & \textbf{Radar Identification}\\ 
\hline\hline
RadYOLOLet~\cite{10804612} & $\geq$ $16$\,dB & $100\%$  & $5$ Types  \\
Waldo~\cite{10001638}      & $\geq$ $17$\,dB & $100\%$ & $1$ Type \\
DeepRadar~\cite{DeepRadar} & $\geq$ $20$\,dB & $99\%$     & $5$ Types \\
Spec-SCAN~\cite{10976013}  & $\geq$ $15$\,dB & $100\%$   & $1$ Type  \\
senseORAN~\cite{senseORAN} & $\geq$ $12$\,dB  & $100\%$ & $1$ Type \\
\hline\textbf{Ours} ({\tt Pipeline 2})        & \textbf{$\geq$ $-5$\,dB} & \textbf{$99.5\%$}& \textbf{$6$ Types}\\
\hline
\end{tabular}
\vspace{-20pt}
}
\end{table}
\vspace{-10pt}
\subsection{Experimental Results on Radar Waveform Identification}
\begin{figure}[!hbt]
    \centering
    \vspace{-14pt}
    \includegraphics[width=0.9\linewidth]{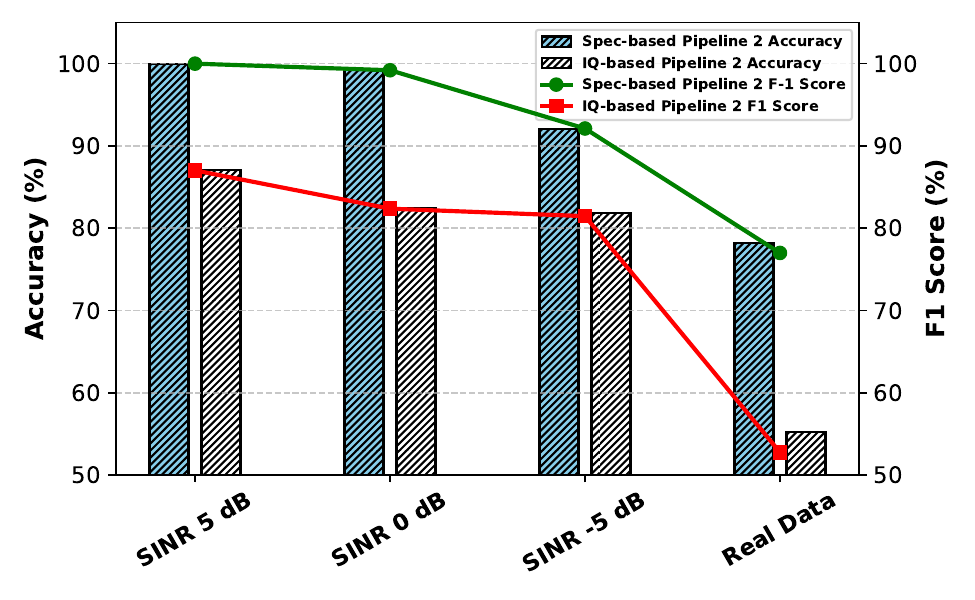}
    \vspace{-8pt}
    \caption{Performance comparison of {\tt Pipeline 1} and {\tt Pipeline 2} for radar waveform identification.}
    \label{fig:accuracy_f1_comparison}
    \vspace{-10pt}
\end{figure}

Figure~\ref{fig:accuracy_f1_comparison} illustrates the comparative performance of spectrogram-based {\tt Pipeline 2} and IQ-based {\tt Pipeline 1}  in terms of accuracy and F$1$-score across varying SINR levels and real testbed data. It is evident that {\tt Pipeline 2} maintain consistently high performance across all conditions, including severe interference at SINR $=-5$\,dB and real-world measurements. This resilience can be attributed to the time-frequency representation provided by spectrograms, which enables the model to capture both transient and stationary signal features essential for distinguishing radar signal characteristics. Notably, the spectrogram model achieves approximately $78\%$ accuracy on real data and exceeds $99\%$ in accuracy, indicating strong discriminability even under domain shift and practical noise.
Conversely, IQ-based models exhibit a steeper decline in both accuracy and F$1$-score, especially at lower SINR values and under real-world conditions. This degradation highlights the limitations of using raw in-phase and quadrature components, which lack explicit temporal and spectral structure and thus fail to expose the discriminative properties required for robust radar signal classification. Specifically, in radar parameter identification tasks involving pulse width, chirp bandwidth, and PRF variations, the absence of structured feature representations in IQ data results in lower model confidence and generalization. This reinforces the necessity of time-frequency domain representations in spectrum sensing applications involving heterogeneous radar signals. 
\vspace{-5pt}

\begin{observation}
We observed that {\tt Pipeline 2} maintain $>99\%$ accuracy for synthetic data with SINR $\geq$ 0\,dB, while {\tt Pipeline 1} reaches maximum accuracy of 88\% for  radar waveform identification for synthetic data with SINR= 5\,dB (see Fig.~\ref{fig:accuracy_f1_comparison}).  
\end{observation}

%% file: sections/Conclusion.tex
\vspace{-0.2cm}
\section{Conclusions}
\vspace{-0.1cm}

\label{sec:discussion_and_conclusion}

In this work, we propose a robust ML-based pipeline for radar detection and waveform identification within the CBRS band. Unlike traditional energy detection methods that degrade under low SINR conditions, our approach leverages high-resolution spectrogram representations to effectively detect radar signals even at SINR levels as low as $-5$ dB, significantly exceeding the FCC-mandated threshold of $99\%$ detection accuracy threshold set at $20$ dB. This enables the commercial $5$G users to transmit at high as $-84$ dBm/MHz when the co-existing radar power is $-94$ dBm/MHz and still able to detect radar presence with $99\%$ accuracy. Furthermore, the framework extends beyond detection of radar by incorporating fine-grained identification of six types of radar waveforms, allowing more context-aware spectrum management. The proposed ML-based pipelines are validated on diverse datasets to demonstrate improvement over state-of-the-art methods.  Future scopes include the radar parameter estimation and distributed execution of the proposed pipelines. The authors
have provided public access to their code and data at~\cite{TWISTLab_RadarPushingBound_GitHub_2025}.

\vspace{-0.4cm}

\section*{Acknowledgment}
\vspace{-0.1cm}
\label{sec:ack}
The authors gracefully acknowledge the sponsorship of the SDR devices from Dr. Stephen Hary through the Airforce Research Lab SDR Challenge program and funding from the US National Science Foundation (CNS 2526490).
\vspace{-4pt}